\documentclass[12pt]{article}

\usepackage[french]{babel}
\usepackage{amssymb}
\usepackage{amsthm}

\usepackage[latin1]{inputenc}

\usepackage{logique}

\author{Jean-Louis Krivine \& Yves Legrandgérard\\
\footnotesize{Paris VII University, C.N.R.S.}\\
}

\title{Valid formulas, games and network protocols} 

\date{\footnotesize {November 14, 2007}}

\begin{document}
\maketitle\noindent

\subsection*{Introduction}
We describe a remarkable relation between a fundamental notion of mathematical logic --
that is {\em valid formula of predicate calculus} -- and the specification of network protocols.
We explain here in detail several simple examples~: the acknowledgement of one or two packets,
and then of an arbitrary number. We show that, using this method, it is possible to specify
the composition of protocols.

We tried to write a self-contained paper, as far as possible, in what concerns the basic notions
of the calculus of predicates. In particular, the notion of {\em valid formula} is defined with the help
of the tools introduced in the present paper (specifically, the game associated with the formula).
The equivalence with the usual definition of this notion in logic is explained in the appendix, but is never used in
the paper.

\subsection*{Logical framework}
The language we use is described below. It is the well known {\em predicate calculus}, fundamental in
mathematical logic~; important restriction~: the only allowed logical symbols are $\to,\bot,\pt$, respectively read as ``~implies~'', ``~false~'', ``~for all~''. In fact, every other logical symbol can be defined with them (see below). This restriction is therefore only syntactic, but not semantic.

\smallskip
We suppose given an infinite set of {\em variables}~: $\{x,y,\ldots\}$, an infinite set of {\em constants}~:
${\cal C}=\{a,b,\ldots\}$ and some {\em predicate symbols} $P\,,Q,R\,,\ldots$~; each of them has an
\emph{arity} which is an integer $\ge0$.

\smallskip\noindent
{\em Atomic formulas} are of the form $\bot$ (read {\em false}) or $Pt_1\ldots t_k$ (denoted also as $P(t_1,\ldots,t_k$)) where $P$ is a predicate symbol of arity $k$ and $t_1,\ldots,t_k$ are variables or constants.\\
Formulas of the predicate calculus are built with the following rules~:

\smallskip\noindent
$\bullet$ An atomic formula is a formula.\\
$\bullet$ If $F$ and $G$ are formulas, then $F\to G$ is a formula (read \gmg $F$ implies $G$\gmd).\\
$\bullet$ If $F$ is a formula and $x$ is a variable, then $\pt x\,F$ is a formula (read
``~for all~$x$,~$F$~'').

\smallskip
{\bfseries Remark.} {\em Propositional calculus} is contained in predicate calculus~: it has the only
logical symbols $\to$ and $\bot$, and only predicate symbols of arity~$0$, usually called {\em propositional variables}.

\smallskip\noindent
We shall systematically use the notation $A,B\to C$ for $A\to(B\to C)$ and, more generally
$A_1,A_2\ldots,A_n\to B$ for $A_1\to(A_2\to(\cdots(A_n\to B)\cdots))$.\\
Usual connectives $\neg,\land,\lor,\dbfl$ of propositional calculus are considered as abbreviations,
and defined as follows~:\\
$\neg F$ is $F\to\bot$~; \ $F\land G$ is $(F\,,G\to\bot)\to\bot$~; \ $F\lor G$ is $\neg F\,,\neg G\to\bot$~;\\
$F\dbfl G$  is $(F\to G)\land(G\to F)$ that is $((F\to G),(G\to F)\to\bot)\to\bot$.

\smallskip
{\bfseries Remark.} The connective {\tt\small XOR}, usual in computer science and often denoted as \ $F\verb?^?\!G$,
can be defined as \ $\neg F\dbfl G$. This abbreviation is not used in the formulas of predicate calculus.

\smallskip
The existential quantifier $\ex$ (read ``~there exists~'') is also considered as an abbreviation~: \ $\ex x\,F$ is
defined as \ $\neg\pt x\neg\,F$ that is \ $\pt x(F\to\bot)\to\bot$.

\smallskip
The notation $\vec{x}$ will denote a finite sequence of variables $x_1,\ldots,x_n$.\\
Therefore, we shall write $\pt\vec{x}$ for $\pt x_1\ldots\pt x_n$ and the same with $\ex$.

\smallskip
In a formula such as $\pt x\,A$, the subformula $A$ is called the {\em scope} of the quantifier~$\pt x$.
An {\em occurrence} of a variable $x$ in a formula $F$ is called {\em bounded} if it is in the scope of a
quantifier $\pt x$~; otherwise, this occurrence is called {\em free}. Given a bounded occurrence of $x$,
the quantifier which bounds it, is by definition, the nearest quantifier $\pt x$ which has this occurrence
in its scope.\\
For instance, in the formula \ $\pt x[\pt x(Rx\to Ry)\to\pt y(Ry\to Rx)]$ there are a bound and a free occurrence of the variable $y$ and two bounded occurrences of the variable $x$. These two occurrences of $x$ are not bounded by the same quantifier.\\
A variable $x$ is called {\em free} in the formula $F$ if there is at least one free occurrence of $x$. The formula
$F$ is called {\em closed} if it contains no free variable.\\
The notation $F[x_1,\ldots,x_n]$ (or $F[\vec{x}]$) will mean that the free variables of the formula $F$ are
 {\em amongst} $x_1,\ldots,x_n$. Then, the formula $\pt x_1\ldots\pt x_n\,F[x_1,\ldots,x_n]$ (or $\pt\vec{x}\,F[\vec{x}]$) is closed.

\smallskip
In any formula $F$, we can rename the {\em bounded} variables in an arbitrary way, provided that no {\em capture of variable} occurs. This means that no free occurrence becomes bound~;
and that any bound occurrence must remain bounded {\em by the same quantifier}. Any formula $G$, obtained from $F$
in this way is considered as identical with $F$.\\
For instance, $\pt z[\pt y(Rx\to Ry)\to Rz]$ is identified with $\pt y[\pt y(Rx\to Ry)\to Ry]$.

\smallskip
For any formula $F[x_1,\ldots,x_k]\equiv F[\vec{x}]$ and constants $a_1,\,\ldots,a_k$, we denote by\\
$F[a_1,\ldots,a_k]\equiv F[\vec{a}]$ the closed formula we obtain by replacing each {\em free}
occurrence of $x_i$ with $a_i$ $(1\le i\le k)$.

\smallskip
{\bfseries Remark.} Any {\em atomic closed} formula $\not\equiv\bot$ has the form $Pa_1\ldots a_k$, where $P$ is a predicate symbol of arity $k$ and $a_1,\ldots,a_k$ are constants. In the interpretation in terms of network protocols which is given below, such a formula represents a {\em packet}, the predicate symbol~$P$ represents the {\em datas}
and $a_1,\ldots,a_k$ represent the {\em header fields} of the packet. When $k=0$, i.e. when $P$ is a
propositional variable, $P$ represents a pure data packet.

\subsubsection*{Normal form of a formula}
A formula is said to be {\em in normal form} or {\em normal}, if it can be obtained by means of the following rules~:\\
$\bullet$ an atomic formula $A$ is normal~;\\
$\bullet$ if $\Phi_1,\ldots,\Phi_n$ are normal, if $A$ is atomic and if $\vec{x}=(x_1,\ldots,x_k)$ is a finite
sequence of variables, then $\pt\vec{x}(\Phi_1,\ldots,\Phi_n\to A)$ is a normal formula.\\
If $n=0$, by definition, this formula is $\pt\vec{x}\,A$.\\
In the same way, if $k=0$ this formula is $\Phi_1,\ldots,\Phi_n\to A$.

\smallskip\noindent
For instance, if $R$ is a unary predicate symbol, the formula $\pt x\,Rx\to\pt x\,Rx$ is not a normal form~;
but $\pt y(\pt x\,Rx\to Ry)$ is a normal form.

\smallskip\noindent
With any formula $F$, we associate its {normal form} $\wh{F}$, which is  obtained as follows~:\\
$\bullet$ if $F$ is atomic, $\wh{F}\equiv F$~;\\
$\bullet$ if $F$ is $\pt x\,G$, then $\wh{F}$ is $\pt x\,\wh{G}$~;\\
$\bullet$ if $F$ is $G\to H$, we write $\wh{H}\equiv\pt\vec{x}(\Phi_1,\ldots,\Phi_n\to A)$. We first rename the (bounded) variables $\vec{x}$ so that they become not free in $G$ (a good method is to use variables that do not appear in $G$)~; then, $\wh{F}$ is $\pt\vec{x}[\wh{G},\Phi_1,\ldots,\Phi_n\to A]$.

\smallskip
{\bfseries Remark.} Obviously, $F$ and $\wh{F}$ have the same free variables. In particular, if $F$ is closed,
then $\wh{F}$ is also closed.\\
Note that any formula of the propositional calculus is in normal form. 

\smallskip\noindent
For instance, the normal form of the formula $(Rx\to\pt x\,Rx)\to\pt x\,Rx$ is~:\\
$\pt z[\pt y(Rx\to Ry)\to Rz]$ or $\pt y[\pt y(Rx\to Ry)\to Ry]$.

\subsection*{The game associated with a closed formula}
Given a closed formula $F$, we define a two players' game~; the players will be called \ {\em $\ex$loise} and {\em$\pt$belard} or, more briefly, $\ex$ and $\pt$ (the same notation as the quantifiers, but no confusion is possible). $\ex$loise is also called the ``~player~'' or the ``~defender~'' and $\pt$belard is called the ``~opponent~''.\\
Intuitively, the player $\ex$ defends the formula $F$, i.e. pretends this formula is ``~true~'' and the opponent
$\pt$ attacks it, i.e. pretends it is ``~false~''.\\
Be careful, there is no symmetry between the players, as it will be seen by the rule of the game. To make the intuitive idea more precise, we can say that $\ex$ pretends the formula $F$ is ``~always true~'' and that $\pt$ pretends it is ``~sometimes false~''.

\smallskip\noindent
Now, we assume that the closed formula $F$ has been put in normal form.\\
Here is the rule of the game associated with this formula [jlk]~:

\smallskip\noindent
We have three finite sets of normal closed formulas, denoted by ${\cal U},{\cal V},{\cal A}$, which change during the play. The elements of the set ${\cal A}$ are closed {\em atomic} formulas. The sets ${\cal U}$ and ${\cal A}$ {\em increase} during the play. At the beginning of the play, we have ${\cal U}=\{F\to\bot\}$, ${\cal V}=\{F\}$
and ${\cal A}=\{\bot\}$ (one-element sets). The first move is done by the opponent $\pt$.\\
Consider now, during the play, a moment when the opponent $\pt$ must play.\\
If, at this moment, the set ${\cal V}$ is empty, the game stops and $\pt$ has lost.\\
Otherwise, he chooses a formula \ $\Phi\equiv\pt\vec{x}(\Psi_1[\vec{x}],\ldots,\Psi_m[\vec{x}]\to A[\vec{x}])$
which is in ${\cal V}$ and a sequence $\vec{a}$ of constants, of the same length as $\vec{x}$.\\
Then {\em he adds} the formulas $\Psi_1[\vec{a}],\ldots,\Psi_m[\vec{a}]$ to the set ${\cal U}$
and also the atomic formula $A[\vec{a}]$ to the set ${\cal A}$. Then the defender $\ex$ must play.\\
She chooses, in the set ${\cal U}$, a formula \
$\Psi\equiv\pt\vec{y}(\Phi_1[\vec{y}],\ldots,\Phi_n[\vec{y}]\to B[\vec{y}])$~; she chooses also a
sequence $\vec{b}$ of constants, of the same length as $\vec{y}$, {\em in such a way that $B[\vec{b}]\in{\cal A}$}~;
this is always possible, since she can, at least, choose $F\to\bot$ which is in ${\cal U}$.\\
Then, {\em she replaces} the content of the set ${\cal V}$ with $\{\Phi_1[\vec{b}],\ldots,\Phi_n[\vec{b}]\}$.\\
Then $\pt$ must play, and so on.

\smallskip
{\bfseries Remarks.} We observe that the opponent $\pt$ wins if, and only if, the play is infinite.
 The play ends after a finite time if, and only if, ${\cal V}$ becomes empty (and then, the player $\ex$ wins).
Just before, the player $\ex$ has chosen an atomic formula which is in ${\cal U}\cap{\cal A}$.\\
The intuitive meaning of the rule of this game is as follows~: at each moment, the defender $\ex$ pretends that one of the formulas of ${\cal U}$ is false and that every formula of ${\cal V}$ is true. On the other hand, the opponent $\pt$ pretends that every formula of ${\cal U}$ is true and that one of the formulas of ${\cal V}$ is false. Now, both agree on the fact that every formula of ${\cal A}$ is false.

In the examples below, we shall interpret a play of this game as a session of communication following a certain protocol. In this interpretation, the opponent $\pt$ is the {\em sender} and the defender $\ex$ is the {\em receiver}.

The disymmetry of the game is well expressed by a celebrated sentence of Jon Postel (known as ``~Postel's law~'' [jp])~: ``~Be conservative in what you send, be liberal in what you receive~''.

\subsection*{Examples}\noindent
1) $F\equiv P\to P$. We can describe an arbitrary play by the following table~:

\smallskip
$\dsp \begin{array}{ccccl}
{\cal U} & {\cal V} & {\cal A} & & \\
(P\to P)\to\bot & P\to P & \bot & &\pt\mbox{ has no choice }\\
(P\to P)\to\bot,P &\hspace{2em}\mbox{unchanged}\hspace{2em} &\bot,P & &\ex\mbox{ chooses }(P\to P)\to\bot\\
\mbox{unchanged} &\mbox{unchanged}&\mbox{unchanged} & &\pt\mbox{ has no choice }\\
\mbox{unchanged}&\mbox{unchanged} &\mbox{unchanged}& &\ex\mbox{ chooses }(P\to P)\to\bot\\
\vdots &\vdots &\vdots & &\hspace{2em}\vdots\\
\mbox{unchanged}&\mbox{unchanged} &\mbox{unchanged}& &\ex\mbox{ chooses }P\\
\mbox{unchanged}&\vide &\mbox{unchanged}& & \ex\mbox{ wins}
\end{array}$

\smallskip\noindent
The different possible plays depend only on the number $n$ of times when $\ex$ chooses the formula $(P\to P)\to\bot$.
If $n$ is infinite, $\ex$ loses.

\smallskip\noindent
2) $F\equiv P\to Q$

\smallskip
$\dsp \begin{array}{ccccl}
{\cal U} & {\cal V} & {\cal A} & & \\
(P\to Q)\to\bot & P\to Q & \bot & &\pt\mbox{ has no choice }\\
(P\to Q)\to\bot,P &\hspace{2em}\mbox{unchanged}\hspace{2em} &\bot,Q & &\ex\mbox{ cannot choose } P\\
\mbox{unchanged} &\mbox{unchanged}&\mbox{unchanged} & &\pt\mbox{ has no choice }\\
\mbox{unchanged}&\mbox{unchanged} &\mbox{unchanged}& &\ex\mbox{ cannot choose} P\\
\vdots &\vdots &\vdots & &\hspace{2em}\vdots
\end{array}$

\smallskip\noindent
There is only one possible play and $\pt$ wins, since this play is infinite.

\smallskip\noindent
3) The reader is invited to study by himself the following two examples~:\\
$((Q\to Q)\to P)\to P$~;  $((P\to Q)\to P)\to P$
(Peirce's law).

\smallskip\noindent
4) $F\equiv\pt x\,Px\to\pt x\,Px$. The normal form of $F$ est $G\equiv\pt y(\pt x\,Px\to Py)$.

\smallskip
$\dsp \begin{array}{ccccl}
{\cal U} & {\cal V} & {\cal A} & & \\
\neg G & G & \bot & &\pt\mbox{ chooses } b_0\\
\neg G,\pt x\,Px &\hspace{2em}\mbox{unchanged}\hspace{2em} &\bot,Pb_0 & &\ex\mbox{ chooses }\neg G\\
\mbox{unchanged} &\mbox{unchanged}&\mbox{unchanged} & &\pt\mbox{ chooses }b_1\\
\mbox{unchanged}&\mbox{unchanged} &\bot,Pb_0,Pb_1& &\ex\mbox{ chooses }\neg F\\
\vdots &\vdots &\vdots & &\hspace{2em}\vdots\\
\mbox{unchanged}&\mbox{unchanged} &\bot,Pb_0,\ldots,Pb_n& &\ex\mbox{ chooses }\pt x\,Px\mbox{ and }b_i\\
\mbox{unchanged}&\vide &\mbox{unchanged}& & \ex\mbox{ wins}
\end{array}$

\smallskip\noindent
Like in example~1, the play only depends on the moment when the player $\ex$ chooses the formula
$\pt x\,Px$ and one of the $b_i$'s already chosen by the opponent $\pt$.

We can give the following interpretation, in terms of network~: the player $\pt$ sends the data packet $P$
with the headers $b_0$, then $b_1$, \ldots\ The acknowledgement by the receiver $\ex$ only happens at the $n$-th step, and it is the packet $Pb_i$ that is acknowledged. Then the play, that is to say the session, stops immediately. From the network point of view, this means that the acknowledgement cannot be lost~; in other words, that the channel from the receiver $\ex$ to the sender $\pt$ is reliable.\\
In the following section, we treat a particularly important example~: the acknowledgement of a packet in a channel which is not reliable.

\subsection*{The formula $\ex x(Px\to\pt y\,Py)$}
Let us call $F$ the normal form of this formula, i.e. $\pt x(\pt y(Px\to Py)\to\bot)\to\bot$. For the sake of clarity, let us put \ $G[x]\equiv\pt y(Px\to Py)$~; thus, we have $F\equiv\neg\pt x\neg\,G[x]$.

The tables I and II below represent what happens during a play, in the (very particular) case when $\ex$
plays in such a way as to win as quickly as possible. There are two possibilities, following what the opponent $\pt$
plays at line~3~:

\bigskip
Table I

\noindent
$\dsp \begin{array}{cccccl}
 & {\cal U} & {\cal V} & {\cal A} & & \\
1 &\neg F & F & \bot & &\pt\mbox{ chooses }F\\
2 &\neg F,\pt x\,\neg G[x] &\hspace{2em}\mbox{unchanged}\hspace{2em} &\mbox{unchanged}& &\ex\mbox{ chooses }\pt x\,\neg G[x]\mbox{ and }a\\
3 &\mbox{unchanged} &G[a]\equiv\pt y(Pa\to Py)&\mbox{unchanged} & &\pt\mbox{ chooses }b\mbox{, with }b\ne a\\
4 &\neg F,\pt x\,\neg G[x],Pa & \mbox{unchanged} &\bot,Pb& &\ex\mbox{ chooses }\pt x\neg G[x]\mbox{ and }b\\
5 &\mbox{unchanged} & G[b]\equiv\pt y(Pb\to Py) &\mbox{unchanged} & &\pt\mbox{ chooses }c\\
6 &\neg F,\pt x\,\neg G[x],Pa,Pb &\mbox{unchanged} &\bot,Pb,Pc & &\ex\mbox{ chooses }Pb\\
7 &\mbox{unchanged}&\vide &\mbox{unchanged}& & \ex\mbox{ wins}
\end{array}$

\bigskip
Table II

\noindent
$\dsp \begin{array}{cccccl}
 & {\cal U} & {\cal V} & {\cal A} & & \\
1 &\neg F & F & \bot & &\pt\mbox{ chooses }F\\
2 &\neg F,\pt x\,\neg G[x] &\hspace{2em}\mbox{unchanged}\hspace{2em} &\mbox{unchanged}& &\ex\mbox{ chooses }\pt x\,\neg G[x]\mbox{ and }a\\
3 &\mbox{unchanged} &G[a]\equiv\pt y(Pa\to Py)&\mbox{unchanged} & &\pt\mbox{ chooses }a\\
4 &\neg F,\pt x\,\neg G[x],Pa & \mbox{unchanged} &\bot,Pa& &\ex\mbox{ chooses }Pa\\
5 &\mbox{unchanged}&\vide &\mbox{unchanged}& & \ex\mbox{ wins}
\end{array}$

\bigskip
But this is only a particular case. The game we are considering presents, in fact, a great variety of possible plays. We shall see that these various plays correspond exactly to the various possibilities which may happen during the
acknowledgement of a packet.\\
The play which is described in table~I represents the case when the communication occurred in the best possible way. We can interpret it as follows~: the receiver $\ex$ begins the session by sending the header $a$ (line~3)~; then the sender $\pt$ sends the packet $Pb$ (line~4)~; $\ex$ receives the packet $Pb$ and sends the acknowledgement (line~5)~; then $\pt$ correctly receives this acknowledgement and sends a signal $Pc$ to terminate the session (line~6).

\smallskip\noindent
Several variants are possible~:\\
i) The player $\ex$ can, at each moment, choose the formula $\neg F$. This corresponds to a re\-initialisation of the session.\\
ii) She can also choose the formula $\pt x\neg\,G[x]$ with an arbitrary header $a'$, which corresponds to no acknowledgement. Then, the opponent $\pt$ must send the packet again. This situation corresponds to the loss of the acknowledgement.\\
iii) In this case, the sender $\pt$ has the possibility of sending $Pa'$ again, which gives to the receiver $\ex$ the possibility of finishing the session immediately by choosing precisely the formula $Pa'$ (since it is now both in ${\cal U}$ and ${\cal A}$).
This corresponds to the case when the sender asks to finish the session. This may happen at the very beginning~: it is the case in the play which is described in table~II  (line~3~: $\pt$ chooses $a$)~; this corresponds to a refusal of opening the session~; then $\ex$ can only close the session, by choosing $Pa$ (again, it is what happens in table~II) or to re-initialise it (by choosing $\neg F$ or $\pt x\neg\,G[x]$).\\
iv) The player $\ex$ can terminate the play by choosing the formula $Pb$, where $b$ is any of the headers sent by $\pt$. This corresponds to a successfull communication session, perhaps after some loss of acknowledgements.

\smallskip\noindent
Any session is a combination of an arbitrary number of such variants.

\subsection*{Sending several packets}
We consider now the case of the acknowledgement of a fixed number $n$ of packets, $n$ being a previously given integer~; the order of the packets must be preserved. The associated formula $F_n$ is defined by recurrence~:\\
$F_1\equiv\ex x\pt y(P_1x\to P_1y)$~; \ $F_{n+1}\equiv\ex x\pt y((F_n\to P_{n+1}x)\to P_{n+1}y)$~; \ $F_n$ is in
normal form.

\smallskip\noindent
For the sake of simplicity, we consider only the case $n=2$. We have the formula~:\\
$F'\equiv\ex x\pt y((F\to Px)\to Py)$ with $F\equiv\ex x\pt y(Qx\to Qy)$.\\
We put $G[x]\equiv\pt y((F\to Px)\to Py)$, $H[x]\equiv\pt y(Qx\to Qy)$~;\\
thus, we have $F'\equiv\pt x\neg G[x]\to\bot$ and $F\equiv\pt x\neg H[x]\to\bot$.\\
The table below describes once more what happens during a play where $\ex$ finishes in the quickest possible way.
For the sake of clarity, in the columns ${\cal U}$ and ${\cal A}$, we shall put, at each line, only the {\em new} formulas.

\bigskip\noindent
$\dsp \begin{array}{rccccl}
 & {\cal U} & {\cal V} & {\cal A} & & \\
1 &\neg F' & F' & \bot & &\pt\mbox{ has no choice}\\
2 &\pt x\,\neg G[x] &\hspace{2em}\mbox{unchanged}\hspace{2em} &\mbox{unchanged}& &\ex\mbox{ chooses }\pt x\,\neg G[x]\mbox{ and }a\\
3 &\mbox{unchanged} &G[a]\equiv\pt y((F\to Pa)\to Py)&\mbox{unchanged} & &\pt\mbox{ chooses }b\mbox{, with }b\ne a\\
4 &F\to Pa & \mbox{unchanged} & Pb& &\ex\mbox{ chooses }\pt x\neg G[x]\mbox{ and }b\\
5 &\mbox{unchanged} & G[b]\equiv\pt y((F\to Pb)\to Py) &\mbox{unchanged} & &\pt\mbox{ chooses }c\\
6 &F\to Pb &\mbox{unchanged} & Pc & &\ex\mbox{ chooses }Pb\\
7 &\mbox{unchanged}& F &\mbox{unchanged}& &\pt\mbox{ has no choice}\\
\multicolumn{6}{c}{\dotfill\mbox{\small\it \ acknowledgement of the first packet\ }\dotfill}\\
8 &\pt x\,\neg H[x] &\hspace{2em}\mbox{unchanged}\hspace{2em} &\mbox{unchanged}& &\ex\mbox{ chooses }\pt x\,\neg H[x]\mbox{ and }d\\
9 &\mbox{unchanged} &H[d]\equiv\pt y(Qd\to Qy)&\mbox{unchanged} & &\pt\mbox{ chooses }e\mbox{, with }e\ne d\\
10 &Qd & \mbox{unchanged} & Qe& &\ex\mbox{ chooses }\pt x\neg H[x]\mbox{ and }e\\
11 &\mbox{unchanged} & H[e]\equiv\pt y(Qe\to Qy) &\mbox{unchanged} & &\pt\mbox{ chooses }f\\
12 &Qe &\mbox{unchanged} & Qf & &\ex\mbox{ chooses }Qe\\
13 &\mbox{unchanged}& \vide &\mbox{unchanged}& &\ex\mbox{ wins}\\
\multicolumn{6}{c}{\dotfill\mbox{\small\it \ acknowledgement of the second packet\ }\dotfill}
\end{array}$

\bigskip\noindent
In this particular case, we essentially get twice the table~I of the previous example. Of course, we may get all the variants already described. But new variants may appear~: indeed, after the acknowledgement of the first packet (lines~8, 10 and~12), the player $\ex$ can, for instance, come back to line~4, that is to say ask again for the first  packet. Thus the receiver may lose a packet, even after having correctly acknowledged it. It is interesting to notice that she has not to acknowledge it again.

\subsection*{Strategies and valid formulas}
Let us consider the game associated with a normal closed formula $F$. A {\em strategy} for $\ex$ in this game is, by definition, a function ${\cal S}$, which takes as an argument a finite sequence of triples
$({\cal U}_i,{\cal V}_i,{\cal A}_i)_{0\le i\le n}$ (${\cal U}_i,{\cal V}_i$ are finite sets of normal closed formulas and ${\cal A}_i$ is a finite set of atomic closed formulas) and gives as a result an ordered pair $(\Psi,\vec{b})$ with $\Psi\in{\cal U}_n$, $\Psi\equiv\pt\vec{y}(\Phi_1[\vec{y}],\ldots,\Phi_k[\vec{y}]\to B[\vec{y}])$, $\vec{b}$ has the same length as $\vec{y}$ and $B[\vec{b}]\in{\cal A}_n$.\\
Intuitively, a strategy ${\cal S}$ for $\ex$loise is a general method which, each time she must play, chooses for
her a possible play, given all the moves already played.\\
The strategy ${\cal S}$ is called a {\em winning strategy} if $\ex$ wins every play following this strategy, whatever
be the choices of $\pt$.\\
We could define in the same way the winning strategies for $\pt$.

\smallskip
A normal closed formula $F$ is called {\em valid} if there exists a winning strategy for $\ex$, in the game associated
with $F$.
Valid formulas are exactly those which correspond to network protocols.

\smallskip\noindent
{\bfseries Games associated with a conjunction or a disjunction.}\\
Given two formulas $F,G$, the game which is associated with the formula $F\land G$, i.e. $(F,G{\to}\bot){\to}\bot$ consists essentially in the following (this is easily checked)~:\\
The opponent $\pt$ chooses one of these two formulas and the game goes on, following the chosen formula~; however,
the player $\ex$ can, at every moment, decide to start again the play from the beginning.\\
With the formula $F\lor G$, it is the player $\ex$ who chooses the formula.

\subsubsection*{Composition of protocols}
Let us consider two valid formulas $F$ and $G$, which correspond respectively to the ``proto\-cols'' (i.e. games)
${\cal P}_F$ and ${\cal P}_G$~; we propose now to build a valid formula $H$ such that the associated protocol ${\cal P}_H$ is~: \ \
${\cal P}_F$ then ${\cal P}_G$.

\smallskip\noindent
Let $A=P(x_1,\ldots,x_n)$ (or $\bot$) be an atomic formula and $F$ a normal formula. An  ``occur\-rence'' of $A$ in $F$
is simply one of the places, in $F$, where $A$ appears.\\
Each occurrence of an atomic formula $A$ in $F$ appears at the end of a subformula of~$F$, of the form
$\pt\vec{x}(\Psi_1,\ldots,\Psi_k\to A)$~; $k$ will be called the \emph{number of hypothesis} of this occurrence of $A$.

\smallskip\noindent
Each occurrence of an atomic formula $A$ in $F$ is either \emph{positive} or \emph{negative}. This property is defined in the following way,
by recurrence on the length of $F$~:

\smallskip\noindent
$\bullet$~~If $F$ is atomic, then $F\equiv A$ and the occurrence of $A$ in $F$ is positive.\\
$\bullet$~~If $F\equiv G\to H$, the occurrence of $A$ in $F$ that we consider, is either in $G$, or in $H$. If it is in $H$,
its sign is the same in $F$ as in $H$. If it is in $G$, it has opposite signs in $F$ and in $G$.\\
$\bullet$~~If $F\equiv\pt x\,G$, the occurrence of $A$ we consider, has the same sign in $F$ and in $G$.

\smallskip\noindent
An atomic occurrence $A$ in $F$, which is \emph{negative and without hypothesis}, will be called a \emph{final atomic occurrence}.
Indeed, it corresponds to the end of a play.

\smallskip\noindent
Now, we can build the formula $H$ we are looking for~: \ \emph{it is obtained by replacing, in~$F$,
each final atomic occurrence $A$ with $G\to A$.}

\smallskip
{\bfseries Remark.} It is easy to show that, if $F$ and $G$ are valid, then the formula $H$  defined in this way is also valid (see
the appendix).

\smallskip
{\bfseries Example.} Take the formula $F\equiv\pt x[\pt y(Px\to Py)\to\bot]\to\bot$ which corresponds to the sending and the
acknowledgement of a packet.\\
Then, we get~: \ $H\equiv\pt x[\pt y((G\to Px)\to Py)\to\bot]\to\bot$.\\
Indeed, there are, in $F$, two atomic negative occurrences, which are $Px$ and the first occurrence of $\bot$.
The only atomic occurrence without hypothesis is $Px$.\\ 
In particular, if we take $G\equiv\pt x[\pt y(Qx\to Qy)\to\bot]\to\bot$, we get the protocol which corresponds to the sending of two packets
(see above).

\subsection*{Formulas and protocols using integer variables}\noindent
We now consider formulas written with a new type of variables~: the ``~integer type~''~; to denote variables of this  type, we shall use the letters $i,j,k,l,m,n$. Thus, there are now two types of variables~: the type
``~integer~'' and the type already defined, which we shall call the type ``~acknowledgement~''~; for the variables of this type, we use as before the letters $x,y,z$.\\
Moreover, we have function symbols {\em on the integer type} (they denote functions from integers to integers), in particular the constant $0$ and the successor $s$ (which represents the function $n\mapsto n+1$). Each function  symbol $f$ has an arity $k\in\ennl$ and represents a well determined function from $\ennl^k$ to $\ennl$ which is
also denoted by $f$. We define the {\em terms} of integer type by the following rules~:

\smallskip\noindent
$\bullet$ an integer variable or a function symbol of arity~$0$ (integer constant) is a term of integer type.\\
$\bullet$ if $f$ is a function symbol of arity $k$ and $t_1,\ldots,t_k$ are terms of integer type, then
$f(t_1,\ldots,t_k)$ is a term of integer type.

\smallskip\noindent
We note that a term of integer type without variable (closed term) represents an integer.

\smallskip\noindent
Predicate symbols are also typed. For example, $Pnx$ or $P(n,x)$ (the first argument of $P$ is of integer type, the second is of type acknowledgement).

\smallskip\noindent
{\bfseries Definition of formulas.}\\
$\bullet$ Atomic formulas ~: $Pt_1\ldots t_k$~; $t_i$ is a constant or a variable of type acknowledgement if
the $i$-th place of $P$ is of this type~; a term of type integer, if the $i$-th place of $P$ is of integer type.\\
$\bullet$ If $F\,,G$ are formulas, $F\to G$ is also a formula.\\
$\bullet$ If $F$ is a formula, $\pt x\,F$ and $\pt n\,F$ are also formulas.\\
$\bullet$ If $F$ is a formula and $t,u$ are terms of integer type, then $t=u\to F$ is also a formula.

\smallskip\noindent
Be careful, the expression $t=u$ alone \emph{is not a formula}.

\smallskip\noindent
{\bfseries  Normal forms.}\\
They are defined as follows~:\\
$\bullet$ An atomic formula is normal.\\
$\bullet$ If $F$ is normal, $\pt x\,F$ and $\pt n\,F$ are normal.\\
$\bullet$ If $A$ is atomic and $\Phi_1,\ldots,\Phi_k$ are normal formulas or expressions of the form $t=u$, then $\Phi_1,\ldots,\Phi_k\to A$ is a normal formula.

\smallskip\noindent
We put the formulas under normal form exactly as in the previous case.

\smallskip\noindent
{\bfseries Game associated with a closed formula under normal form.}\\
We indicate here only the additions to the game rule which has been already given~:

\smallskip\noindent
i) when one of the players has chosen a formula $\pt\vec{\xi}(\Phi_1,\ldots,\Phi_n\to A)$, ($\vec{\xi}=(\xi_1,\ldots,\xi_n)$ where $\xi_i$ is a variable of type integer or acknowledgement)~:

\begin{itemize}
\item first, he or she chooses some values $\vec{a}$ for $\vec{\xi}$.
\item then, he or she computes the (boolean) expressions $\Phi_j$ of the form $t_j=u_j$.
\item if all of them are true, we get rid of them and the play goes on as before with the (simpler) formula obtained in this way.
\item if any of them is false~:

if $\ex$ is playing, she must choose other values $\vec{a}$ or another formula (which is always possible, as we already saw). 

if the opponent $\pt$ is playing, then he has lost and the play stops.
\end{itemize}

\smallskip\noindent
ii) as in the previous game, when the player $\ex$ chooses, in the set ${\cal U}$, a formula~:\\
$\Psi\equiv\pt\vec{y}(\Phi_1[\vec{y}],\ldots,\Phi_n[\vec{y}]\to B[\vec{y}])$ and a sequence $\vec{b}$ of constants,
of the same length as~$\vec{y}$, she have to check that $B[\vec{b}]\in{\cal A}$, i.e. to check that two atomic closed formulas are identical. These formulas may contain closed terms of type integer and \emph{these terms must be
computed before comparing them.}

\smallskip\noindent
{\bfseries $\omega$-valid formulas.}\\
A normal closed formula $F$ will be called {\em $\omega$-valid} if there exists a winning strategy for~$\ex$, in the game associated with $F$. The $\omega$-valid formulas are exactly those which correspond to network protocols (as before, with valid formulas).

\smallskip\noindent
{\bfseries Example.}\\
First, $\pt$ send an integer $n$~; after that, acknowledgement of $n$ packets. The formula is~:\\
$F\equiv\pt j\{\pt i[j=si\to\ex x\pt y((Ui\to Pix)\to Piy)]\to Uj\}\to\pt n\,Un$.\\
$U$ is a predicate symbol with one argument of integer type~; $P$ has two arguments, the first is of integer
type, the second of type acknowledgement.

\smallskip\noindent
Put \ $G\equiv\pt j\{\pt i[j=si\to\ex x\pt y((Ui\to Pix)\to Piy)]\to Uj\}$ and\\
$H[i,x]\equiv\pt y((Ui\to Pix)\to Piy)$.\\
We put the formula $F$ in normal form, so it is written as $F\equiv\pt n(G\to Un)$.

The following table shows the particular session in which every packet is acknowledged in the quickest possible way.

\bigskip
\hspace*{-3.0em}{\small
$\dsp \begin{array}{rccccl}
 & {\cal U} & {\cal V} & {\cal A} & & \\
1 &\neg F & F & \bot & &\pt\mbox{ chooses }n_0\\
2 & G &\mbox{unchanged} & Un_0 & &\ex\mbox{ chooses }G\mbox{ and }n_0\\
3 & \mbox{unchanged} &\hspace{-1em}\pt i(n_0{=}si{\to}\ex xH[i,x]) & \mbox{unchanged} & &\pt\mbox{ can only choose }n_0-1\\
4 &\pt x\neg H[n_0{-}1,x] & \mbox{unchanged} & \mbox{unchanged}& &\ex\mbox{ chooses }a_0\\
5 &\mbox{unchanged} & H[n_0{-}1,a_0] &\mbox{unchanged} & &\pt\mbox{ chooses }b_0\ne a_0\\
6 &U(n_0{-}1){\to} P(n_0{-}1,a_0) &\mbox{unchanged} & P(n_0{-}1,b_0) & &\ex\mbox{ chooses } \pt x\neg H[n_0{-}1,x]\mbox{ and }b_0\\
7 &\mbox{unchanged}& H[n_0{-}1,b_0] &\mbox{unchanged}& &\pt\mbox{ chooses }b_1\ne a_0, b_0\\
8 &U(n_0{-}1){\to} P(n_0{-}1,b_0) &\mbox{unchanged} & P(n_0{-}1,b_1) & &\ex\mbox{ chooses } U(n_0{-}1){\to} P(n_0{-}1,b_0)\\
9 &\mbox{unchanged} & U(n_0{-}1) & \mbox{unchanged} & & \pt\mbox{ can only choose }U(n_0{-}1)\\
\multicolumn{6}{c}{\dotfill\mbox{\small\it \ acknowledgement of packet $n_0{-}1$\ }\dotfill}\\
10 &\mbox{unchanged} & \mbox{unchanged} & U(n_0{-}1) & &\ex\mbox{ chooses }G\mbox{ and }n_0{-}1\\
\vdots & \vdots & \vdots & \vdots & & \hspace{2em}\vdots\\
\multicolumn{6}{c}{\dotfill\mbox{\small\it \ acknowledgement of packet $0$\ }\dotfill}\\
&\mbox{unchanged} & \mbox{unchanged} & U0& &\ex\mbox{ chooses }G\mbox{ and }0\\
&\mbox{unchanged} & \pt i(0{=}si{\to}\ex xH[i,x]) & \mbox{unchanged} & &\pt \mbox{ has lost}\\
\end{array}$}

\bigskip\noindent
In order to avoid a supplementary complication, we did not ask that the integer $n$ (the number of packets to transmit) which is sent by $\pt$, be acknowledged. But if we want this integer to be acknowledged, we must add a
field ``~acknowledgement~'' to the predicate symbol $U$, which therefore becomes binary. In this case, we write the
following formula~:\\
$F\equiv\pt n\ex x'\pt y'((\neg\,G[x']\to Unx')\to Uny')$ with\\
$G[x']\equiv\pt j\{\pt i[j=si\to\ex x\pt y((Uix'\to Pix)\to Piy)]\to Ujx'\}$

\smallskip\noindent
The following table shows the beginning of a communication session and the acknowledgement of the integer $n$.\\
We put $H[n,x']\equiv\pt y'((\neg\,G[x']\to Unx')\to Uny')$.

\bigskip\noindent
$\dsp \begin{array}{cccccl}
 & {\cal U} & {\cal V} & {\cal A} & & \\
1 &\neg F & F & \bot & &\pt\mbox{ chooses }n_0\\
2 &\pt x'\,\neg H[n_0,x'] &\hspace{2em}\mbox{unchanged}\hspace{2em} &\mbox{unchanged}& &\ex\mbox{ chooses }\pt x'\,\neg H[n_0,x']\mbox{ and }x'_0\\
3 &\mbox{unchanged} & H[n_0,x'_0] &\mbox{unchanged} & &\pt\mbox{ chooses }y'_0\\
4 &\neg G[x'_0]\to Un_0x'_0 & \mbox{unchanged} & Un_0y'_0& &\ex\mbox{ chooses }\pt x'\neg H[n_0,x']\mbox{ and }y'_0\\
5 &\mbox{unchanged} & H[n_0,y'_0] &\mbox{unchanged} & &\pt\mbox{ chooses }y'_1\\
6 &\neg G[y'_0]\to Un_0y'_0 & \mbox{unchanged} & Un_0y'_1& &\ex\mbox{ chooses }\neg G[y'_0]\to Un_0y'_0\\
7 &\mbox{unchanged}&\neg\,G[y'_0] &\mbox{unchanged}& & \pt\mbox{ has no choice}\\
8 & G[y'_0]&\mbox{unchanged}&\mbox{unchanged}& &\ex\mbox{ chooses }G[y'_0]\mbox{ and }n_0\\
  &\vdots&\vdots&\vdots & &\hspace{2em}\vdots\\
\end{array}$

\bigskip\noindent
From now on, the play continues as in the previous example (with the formula $G[y'_0]$ instead of the formula $G$).

\smallskip
{\bfseries Remark.} A somewhat simpler formula for the same protocol can be written as~:\\
$\pt n\ex x'\pt y'(G[x']\to Uny')$ with, as before~:\\
$G[x']\equiv\pt j\{\pt i[j=si\to\ex x\pt y((Uix'\to Pix)\to Piy)]\to Ujx'\}$.\\
The reader will check this easily.

\subsection*{Appendix}\noindent
{\bfseries  Valid formulas.}\\
The usual definition of a valid formula of the predicate calculus uses the notion of \emph{model}. The interested reader will find it, for example in~[rcdl] or [jrs]. A formula is called valid if it is satisfied in every model. A fundamental theorem of logic, known as the {\em completeness theorem}, says that a formula is valid if, and only if, it is provable by means of the deduction rules of ``~pure logic~''.\\
This notion of validity is equivalent to that introduced in the present paper, which is given in terms of strategies (see a proof in [jlk]).\\
It is often much easier to check the validity of a formula with the help of models. For example, it is immediately seen in this way that the formula $F\equiv\ex x(Px\to\pt y\,Py)$ is valid~: indeed, either the model we consider satisfies $\pt y\,Py$ and therefore also $F$, either it satisfies $\ex x\neg\,Px$ and thus again $F$.

\smallskip
Let us consider two valid formulas $F$ and $G$, and let $H$ be the formula defined above, such that the protocol ${\cal P}_H$
associated with $H$ is obtained by composing the protocols associated with $F$ and $G$. Then, it is easy to show that $H$ is valid.
Indeed, we obtained the formula $H$ by replacing, in $F$, some subformulas $A$ with $G\to A$. But $A$ and $G\to A$ are obviously
equivalent, since $G$ is valid. Thus, we get finally a formula $H$ wich is equivalent to $F$, and therefore a valid formula.

\smallskip
For the formulas with two types (integer and acknowledgement), the situation is a bit more complex. The $\omega$-valid  formulas are the formulas which are satisfied in every {\em $\omega$-model}, that is to say the models in which the integer type has its standard interpretation. Again, in this case, it is often much easier to use this definition in order to check that a formula is $\omega$-valid.\\
For instance, it is not difficult to show that the formula (that we have already used before) $F\equiv G\to\pt n\,Un$, with \ $G\equiv\pt j\{\pt i[j=si\to\ex x\pt y((Ui\to Pix)\to Piy)]\to Uj\}$ is $\omega$-valid.\\
Indeed, we assume $G$ and we show $Un$ by recurrence on the integer $n$.\\
Proof of $U0$. We put $j=0$ in $G$~; since $0=si$ is false, we get $U0$.\\
Proof of $Un\to U(n+1)$. We put $j=n+1$ in $G$. Then, it suffices to show~:\\
$\pt i[n+1=si\to\ex x\pt y((Ui\to Pix)\to Piy)]$ with $Un$ as an hypothesis. Since $n+1=si$ is equivalent to $i=n$, we now need to show $\ex x\pt y((Un\to Pnx)\to Pny)$, that is to say $\ex x\pt y(Pnx\to Pny)$ (because we assume $Un$). But this last formula is already shown.

\smallskip\noindent
With some simple changes, the same proof works for the formulas~:\\
$\pt n\ex x'\pt y'((\neg\,G[x']\to Unx')\to Uny')$ \ and \ $\pt n\ex x'\pt y'(G[x']\to Uny')$\\
with \ $G[x']\equiv\pt j\{\pt i[j=si\to\ex x\pt y((Uix'\to Pix)\to Piy)]\to Ujx'\}$.\\
Indeed, you only need to show the first one, since the second is trivially weaker.

\smallskip\noindent
{\bfseries Determination of games.}\\
A game is called {\em determined} if one of the players has a winning strategy. It is always the case for the games considered in this paper (Gale-Stewart theorem).\\
Sketch of proof. Suppose that $\ex$ has no winning strategy. Then, the following strategy is winning for the opponent $\pt$~: to play, at each step, in such a way that the player $\ex$ has no winning strategy from this step. Then the play lasts infinitely long, and $\pt$ wins.

\subsection*{References}\noindent
{[}rcdl] René Cori \& Daniel Lascar. {\em Mathematical logic.} Oxford University Press (2001).\\
{[}jlk] Jean-Louis Krivine. {\em Realizability~: a machine for analysis and set theory.} Geocal06, Marseille (2006).\\
{\ttfamily http\hspace{-0.6em}://cel.archives-ouvertes.fr/cel-00154509}. More recent version at~:\\
{\ttfamily http\hspace{-0.6em}://www.pps\hspace{-0.1em}.jussieu.fr/\~{ }krivine/articles/Mathlog07\hspace{-0.3em}.pdf}\\
{[}jp] Jon Postel. {\em RFC 793 - Transmission Control Protocol specification.} (1981).\\
{[}jrs] Joseph R. Schoenfield. {\em Mathematical logic.} Addison Wesley (1967).

\end{document}